\documentclass[
twocolumn,
superscriptaddress,
amsmath,
amssymb,
aps,
pre,
]{revtex4-1}

\usepackage{graphicx}   % Include figure files
\usepackage{dcolumn}    % Align table columns on decimal point
\usepackage{bm}         % bold math
\usepackage{color}      % change color of text
\usepackage[extension=xxx]{hyperref}
\usepackage{physics}
\begin{document}

% TITLE

% 2nd possibility
\title{Asymmetric Heat Transport in Ion Crystals}

%%%%%%%%%%%%%%%%%%%%%%%%%%%%%%%%%%%%%%%%%%%%%%%%%%%%%%%%%%%%%%%%%%%%%%%%%%%%%%%%%%%%%%%%%%%%%%%%%%%%

% AUTHORS
\author{M. A. Sim\'{o}n}
\email[]{miguelangel.simon@ehu.eus}
\affiliation{Departamento de Qu\'{i}mica-F\'{i}sica, Universidad del Pa\'{i}s Vasco, UPV- EHU - Bilbao, Spain}

\author{S. Mart\'inez-Garaot}
\affiliation{Departamento de Qu\'{i}mica-F\'{i}sica, Universidad del Pa\'{i}s Vasco, UPV- EHU - Bilbao, Spain}

\author{M. Pons}
\affiliation{Departamento de F\'isica Aplicada I, Universidad del Pa\'{i}s Vasco, UPV- EHU - Bilbao, Spain}

\author{J. G. Muga.}
\email[]{jg.muga@ehu.es}
\affiliation{Departamento de Qu\'{i}mica-F\'{i}sica, Universidad del Pa\'{i}s Vasco, UPV- EHU - Bilbao, Spain}

% ABSTRACT
\begin{abstract}
We numerically demonstrate heat rectification for linear chains of  ions in trap lattices with  graded  trapping frequencies, in contact with
thermal baths implemented by optical molasses.  To calculate the local temperatures and heat currents we find the stationary state
by solving a system of algebraic equations. This approach is much faster than the usual method
that integrates the dynamical equations of the system and averages over noise realizations.
\end{abstract}

% MAKETITLE COMMAND
\maketitle

% BODY OF THE ARTICLE

\section{Introduction\label{Introduction}}

The ideal thermal rectifier, also ``thermal diode'',  is a device that allows heat to propagate in one direction, from a hot to a cold bath, but not in the opposite one when the temperature bias of the baths is  reversed. The name is set by analogy to the
half-wave rectifiers or diodes for electric current. More generally thermal rectification simply denotes
asymmetric heat flows (not necessarily all or nothing) when the bath temperatures are reversed.
 Thermal rectification was discovered by C. Starr in 1936 in a junction between copper and cuprous oxide \cite{Starr1936}. Many years later, a work of Terraneo \textit{et al.} demonstrated thermal rectification in a model
consisting on a segmented chain of coupled nonlinear oscillators  in contact with two thermal baths at temperatures $T_H$ and $T_C$, with $T_H > T_C$ \cite{Terraneo2002}. This paper sparked a substantial body of research  to this day \cite{Pereira2019} (see Fig. 1 in \cite{Roberts2011}).
%They showed asymmetric heat transport in a chain of coupled oscillators with on-site, non-harmonic and parity-breaking potentials whose two ends are in contact with  The temperature bias between the two baths produces an out-of-equation energy flux from the hot to the cold bath which is significantly reduced when the temperatures of the baths are exchanged.
%
%Their work showed rectification in a model that was usually employed to describe the unfolding dynamics of DNA chains \cite{PhysRevLett.85.6,PhysRevE.47.684} consisting of a chain of coupled oscillators with on-site, non-harmonic and parity-breaking potentials. When the ends of the chain are in contact with heat reservoirs at temperatures $T_{H}$ and $T_{C}$, respectively, an out-of-equilibrium energy current (in form of mechanical oscillation) flows from the hotter one to the colder one. If the temperatures of the baths are exchanged, the heat current from the hot reservoir to the cold one is significantly hindered.

\begin{figure*}
    \centering
    \includegraphics[width=0.9\linewidth]{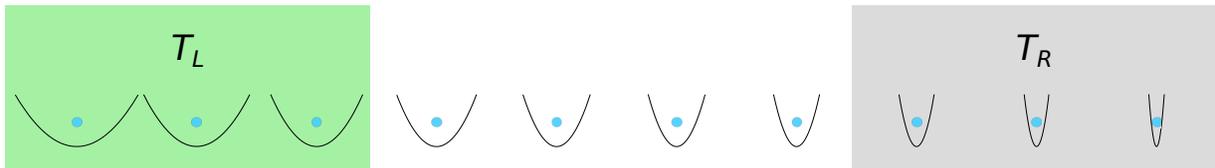}
    \caption{(Color online, Not to scale)  Diagram illustrating the graded chain of trapped ions that is proposed as thermal rectifier in this article. The left and right ends of the chain are in contact with optical molasses at temperatures $T_L$ and $T_R$ (green and grey boxes respectively). Each ion is in an individual trap, the frequency of which is controlled independently from the others. The frequencies of the traps increase homogeneously from left to right, forming a graded structure. The ions interact through the Coulomb force, which is long range, and therefore all the ions interact between them, even distant neighbors. The number of ions in this figure does not correspond to the one used in the calculations.}
    \label{fig:Diagram}
\end{figure*}

Research on thermal rectification has gained a lot of attention in recent years as a key ingredient to build
prospective devices to control heat flows similarly to electrical currents \cite{Roberts2011,Li2012}. There are fascinating proposals
%that exploit the analogy with electronic currents
to engineer thermal logic circuits \cite{Ye2017} in which information, stored in thermal memories \cite{Wang2008}, would be processed in thermal gates \cite{Wang2007}. Such thermal gates, as their electronic counterparts,  will require thermal diodes and thermal transistors  to operate \cite{Casati2006,Joulain2016}.
%Information may also be stored in thermal memories \cite{Wang2008}.
Heat rectifying devices would also be quite useful  in nano electronic circuits, letting delicate components dissipate heat while being protected from external heat sources \cite{Roberts2011}.

Most work on thermal diodes has been theoretical with very few experiments. A relevant attempt to build a
thermal rectifier was based on a graded structure made of carbon and boron nitride nanotubes that transports heat between a pair of heating/sensing circuits \cite{Chang2006}. One of the ends of the nanotube is loaded with a deposition of another material, which makes the heat flow better from the loaded end to the unloaded end. However, rectifications were small, with rectification factors (relative
heat-flow differentials) around $7\%$.
% in \cite{Chang2006}).

Much effort has been aimed at improving the rectification factors and the features of the rectifiers. Some works
relied, as in \cite{Terraneo2002}, on a chain segmented into two or more regions  with different properties,  but using other lattice models such as the Frenkel-Kontorova (FK) model \cite{Li2008,Hu2006}.
The fundamental ingredient for having rectification  was attributed to nonlinear forces in the chain \cite{Zeng2008,Katz2016,Li2008,Hu2006,Benenti2016,Li2012}, which lead to a temperature dependence of the phonon bands or power spectral densities. The bands may match or mismatch at the
interfaces depending on the sign of the temperature bias of the baths, allowing or obstructing heat flow \cite{Terraneo2002,Li2004}.  Later, alternative mechanisms have been
also proposed which do not necessarily rely on anharmonic potentials \cite{Pereira2017,Pons2017}.

It was soon realized that the performance of segmented rectifiers was very sensitive to the size of the device, i.e., rectification decreases with increasing the length of the rectifier \cite{Hu2006}. To overcome this limitation two ideas were proposed. The first one consists in using graded rather than segmented chains, i.e., chains where some physical property varies continuously along the site position such as the mass of particles in the lattice \cite{Wang2012,Chen2015,Romero-Bastida2017,Yang2007,Romero-Bastida2013,Dettori2016,Pereira2010,Pereira2011,Avila2013}. The second one uses particles with long range interactions (LRI), such that all the particles in the lattice interact with each other \cite{Chen2015,Bagchi2017,Pereira2013}. The rationale behind is that in a graded system new asymmetric, rectifying channels are created, while the long range interactions create
also new transport channels, avoiding the usual decay of heat flow with size \cite{Chen2015}.
%
% ``\textit{with long-range interactions (LRI), we conjecture that the presence of new links (interactions) among different sites creates new channels for the heat transport... Besides that, in a graded system... new asymmetric channels are created which in turn favors the asymmetric flow, i.e., rectification}'' \cite{0295-5075-111-3-30004}.
%
Besides a stronger rectification power, LRI graded chains are expected to have better heat conductivity than segmented ones. This is an important point for technological applications, because devices with high rectification factors are not useful if the currents that flow through them are very small.

In this article we propose to bridge the gap between mathematical models and actual systems
exploring the implementation of a heat rectifier in a realistic, graded system with long range interactions:
a chain of  ultracold ions in a segmented Paul trap with graded microtraps for each ion.
Long-range interactions are due to the Coulomb forces, and the  baths at the ends of the chain  may be implemented with optical molasses,
see Fig. \ref{fig:Diagram}. The trapping frequencies of the  microtraps are controlled individually in order to create a graded and asymmetric trap-frequency profile along the chain. This asymmetry will lead to a heat flow that depends on the sign of the temperature difference of the baths. Heat transport in trapped ion chains has been studied in several works  \cite{Freitas2016,Ruiz2014,Pruttivarasin2011,Ramm2014} and interesting phenomena like phase transitions have been investigated \cite{Freitas2016,Ruiz2014,Pruttivarasin2011}. The idea of using locally-controlled traps is already mentioned in \cite{Freitas2016} to implement disorder and study its effects. The device we present here may be challenging to implement, but at reach with the current technology, in particular  that of microfabricated traps \cite{Cirac2000,Krauth2014,Schmied2009}. The setting is thought for a small number of controllable ions, and we refrain from scaling up the
size of the chain to keep the simulations realistic.

The rest of the article is organized as follows. In Section \ref{Physical System} we describe the physical system of trapped ions with graded trap frequencies. We also set the stochastic dynamics due to the action of lasers at the chain edges.
In Section \ref{sec:HeatFlow}  we implement an efficient  method to find the steady state using Novikov's theorem and solving an algebraic system of equations. In Section \ref{Numerical Results} we present simulations of this system exhibiting thermal rectification and discuss the advantages/disadvantages of using a graded frequency profile instead of a segmented one. Finally, in Section \ref{Conclusions} we summarize our conclusions and discuss future research.

\section{Physical System\label{Physical System}}
Consider a linear lattice of $N$ individual harmonic traps of (angular) trapping frequencies  $\omega_n$ evenly distributed along the $x$ axis at a distance $a$ from each other. Each trap contains a single ion that interacts with the rest via Coulomb potentials. All the ions are of the same species, with mass $m$ and charge $q$. The Hamiltonian that describes the dynamics of the system is (we consider only linear, one dimensional motion along the chain axis)
\begin{equation}
    H(\bm{x},\bm{p}) = \sum_{n=1}^N \left[\frac{p_n^2}{2m}  + \frac{m\omega_n^2}{2} (x_n - x_n^{(0)})^2\right] + V_{int}(\bm{x}),
    \label{eq:ChainHamiltonian}
\end{equation}
where $\{x_n,p_n\}$, position and momentum of each ion, are the components of  the vectors
$\bm{x},\bm{p}$, $x_n^{(0)} = n  a$ are the centers of the harmonic traps, and $V_{int}$ is the sum of the Coulomb interaction potential between all  pairs of ions,
\begin{equation}
    V_{int}(\bm{x}) = \frac{1}{2}\sum_n \sum_{l\neq n} V_{C}(\left|x_n-x_l\right|),
    \label{eq:InteractionHamiltonian}
\end{equation}
with $V_{C}(\left|x_n-x_l\right|) = \frac{q^2}{4\pi\varepsilon_0}\frac{1}{\left|x_n-x_l\right|}$. The ends of the chain are in contact with two thermal reservoirs at temperatures $T_L$ for the left bath and $T_R$ for the right bath respectively. The action of the resevoirs on the dynamics of the chain is modeled via Langevin baths at temperatures $T_L$ and $T_R$ \cite{Lepri2003,Dhar2018}. The equations of motion of the chain, taking into account the baths and the Hamiltonian, are
\begin{equation}
    \begin{split}
        \dot{x}_n &= \frac{1}{m}p_n, \\
        \dot{p}_n &= - m\omega_n^2 (x_n-x_n^{(0)}) - \frac{\partial V_{int}}{\partial x_n} - \frac{\gamma_n}{m}p_n + \xi_n(t),
    \end{split}
    \label{eq:Dynamics}
\end{equation}
where $\gamma_n$ and $\xi_n(t)$ are only non-zero for the ions in the end regions, in contact with the left and right baths in the sets $L = \left\{1,2,...,N_L\right\}$ and $R = \left\{N-(N_R-1),...,N-1,N\right\}$,  see Fig. \ref{fig:Diagram}. The $\gamma_n$ are friction coefficients and $\xi_n(t)$ are uncorrelated Gaussian noise forces satisfying $\expval{ \xi_n(t)} = 0$ and $\expval{ \xi_n(t)\xi_m(t') } = 2 D_n \delta_{nm}\delta(t-t')$, $D_n$ being the diffusion coefficients. These Gaussian forces are formally the time derivatives of independent Wiener processes (Brownian motions)   $\xi_n(t) = \sqrt{2D_n}\frac{dW_n}{dt}$ \cite{Toral2014,Ruiz2014} and Eq. \eqref{eq:Dynamics} is a stochastic differential equation (SDE) in the Stratonovich sense \cite{Toral2014}.
%The isolated system of trapped ions would evolve under the Hamilton equations $\dot{x}_n = \partial_{p_n}H= p_n/m$ and $\dot{p}_n = -\partial_{x_n}H$ but the Langevin bath introduces a friction and a noisy term into Eq. \eqref{eq:Dynamics}.

The baths are physically implemented by optical molasses consisting of a pair of counterpropagating Doppler-cooling lasers \cite{Ruiz2014}. The friction and diffusion coefficients for the ions in contact with the baths are given by \cite{Arimondo1993}
\begin{equation}
    \begin{split}
        \gamma_n &= -4 \hbar k_{L,R}^2 \left(\frac{I_{L,R}}{I_0}\right)\frac{2\delta_{L,R}/\Gamma}{\left[1 + (2\delta_{L,R}/\Gamma)^2\right]^2},\\
        D_n &= \hbar^2 k_{L,R}^2 \left(\frac{I_{L,R}}{I_0}\right) \frac{\Gamma}{1 + (2\delta_{L,R}/\Gamma)^2},\\
        n &\in L,R,
    \end{split}
    \label{eq:DopplerCooling}
\end{equation}
where $k_L$ ($k_R$) and $I_L$ ($I_R$) are the wave vector and intensity of the left (right) laser. $\delta_L$ ($\delta_R$) is the detuning of the left (right) laser with respect to the angular frequency $\omega_0$ of the atomic transition the laser is exciting, and $\Gamma$ is the corresponding natural line-width of the  excited state. The expressions in Eq. \eqref{eq:DopplerCooling} are valid only if the intensities of the lasers are small compared to the saturation intensity $I_0$, $I_{L,R}/I_0\ll 1$. In this bath model, the friction term in Eq. \eqref{eq:Dynamics} comes from the cooling action of the laser and the white noise force $\xi_n(t)$ corresponds to the random recoil of the ions due to spontaneous emission of photons \cite{Metcalf1999,Metcalf2003,Cohen1990}. Using the diffusion-dissipation relation $D = \gamma k_B T $ \cite{Chee2010}, the temperature of the optical molasses baths are given by
\begin{equation}
    T_{L,R} = -\frac{\hbar \Gamma}{4 k_B} \frac{1+(2\delta_{L,R}/\Gamma)^2}{(2\delta_{L,R}/\Gamma)},
    \label{eq:DopplerCoolingTemperature}
\end{equation}
with $k_B$ being the Boltzmann constant. If the laser intensities are low enough, the temperatures of the baths are controlled by modifying the detunings. When $\delta = -\Gamma / 2$ the optical molasses reach their minimum temperature possible, the Doppler limit $T_{D} = {\hbar \Gamma}/({2k_B})$.
\section{Calculation of the stationary heat currents\label{sec:HeatFlow}}
The local energy of each site is defined by
\begin{equation}
    H_n = \frac{1}{2m} p_n^2 + \frac{1}{2}m\omega_n^2 \left( x_n - x_n^{(0)}\right)^2 +\frac{1}{2}\sum_{l\neq n} V_C(\left|x_n-x_l\right|).
    \label{eq:LocalEnergy}
\end{equation}
Differentiating $H_n$ with respect to time we find the continuity  equation
\begin{equation}
    \dot{H}_n = \frac{p_n}{m}\! \left[ \xi_n(t)-\gamma_n \frac{p_n}{m} \right]\! - \frac{1}{2m}\!\sum_{l\neq n}\frac{\partial V_C (\left|x_n\!-\!x_l\right|)}{\partial x_n}(p_n + p_l).
    \label{eq:continuityFirst}
\end{equation}
Two different contributions can be distinguished: $j^B_n \equiv \frac{p_n}{m} \left[ \xi_n(t)-\gamma_n \frac{p_n}{m} \right]$, which is the energy flow from the laser reservoir to the ions at the edges of the chain (only for $n\in L,R$), and $\dot{H}_n^{int} \equiv - \frac{1}{2m}\sum_{l\neq n}\frac{\partial V_C (\left|x_n-x_l\right|)}{\partial x_n}(p_n + p_l)$, which gives the ``internal'' energy flow due to the interactions with the rest of the ions. In the steady state $\expval{\dot{H}_n} = 0$, and therefore
\begin{equation}
    \expval{j^B_n} + \expval{\dot{H}_n^{int}} = 0,
    \label{eq:continuitySecond}
\end{equation}
where $\langle \cdot\!\cdot\!\cdot \rangle$ stands for the expectation value with respect to  the ensemble of noise processes $\bm\xi (t)$ ($\bm\xi$ represents a vector with
components $\xi_n$). Equation \eqref{eq:continuitySecond} implies that, in the steady state, the internal rates $\dot{H}_n^{int}$ vanish for the inner ions of the chain because $j^B_n = 0$ for $n\notin L,R$. In chains with nearest-neighbor (NN) interactions,
$\langle\dot{H}_n^{int}\rangle$ simplifies to two compensating and equal-in-magnitude contributions that define the homogeneous heat flux across the chain.
For long-range interactions this is not so and defining the flux is not so straightforward. A formal possibility is to impose
nearest-neighbor interatomic interactions for some atoms in the chain \cite{Chen2015},
but this approach is not realistic in the current system so we define instead the heat currents for the left and right baths as
\begin{equation}
    \begin{split}
        J_L &= \sum_{n\in L} \expval{j^B_n},\\
        J_R &= \sum_{n\in R} \expval{j^B_n},
    \end{split}
    \label{eq:BathHeatFlows}
\end{equation}
respectively. In the steady state we must have $J_L + J_R = 0$, since the local energies stabilize and internal energy
flows cancel. We use either $J_L$ or $J_R$ to calculate the total energy flow in the chain, always taking the absolute value, i.e., $J \equiv \abs{J_L}= \abs{J_R}$. $J$ is defined as $J_\rightarrow$ when the hot bath is on the left
and $J_\leftarrow$ when it is on the right.

To compute the average heat fluxes of the baths $\expval{j^B_n}$ in Eq. \eqref{eq:BathHeatFlows} we need
the averages $\expval{p_n(t)\xi_n(t)}$. Instead of explicitly averaging $p_n(t)\xi_n(t)$ over different realizations of the white noise we use Novikov's theorem \cite{Novikov1965,Ma2011,Toral2014}. Novikov's theorem states that the ensemble average (over  the realizations of the noise) of the product of some functional $\phi(t)$, which depends on a Gaussian noise $\xi(t)$ with zero mean value, $\expval{\xi(t)} = 0$, and the noise itself, is given by
\begin{equation}
    \expval{\xi(t)\phi(t)} = \int_0^t dt' \expval{\xi(t)\xi(t')} \expval{\frac{\delta \phi(t)}{\delta\xi(t')}},
    \label{eq:Novikov_GeneralExpression}
\end{equation}
where ${\delta \phi(t)}/{\delta\xi(t')}$ is the functional derivative of $\phi(t)$ with respect to the noise, with $t'<t$. When the noise is $\delta-$correlated, $\expval{\xi(t)\xi(t')}=2D\delta(t-t')$, and Eq. \eqref{eq:Novikov_GeneralExpression} reads $\expval{\xi(t)\phi(t)} = D \expval{{\delta \phi(t)}/{\delta\xi(t')}}|_{t' \to t^-}$. To apply Novikov's theorem to our model we need the functional derivatives of the position $x_n(t)$ and momentum $p_n(t)$ coordinates with respect to the white noises. We integrate Eq. \eqref{eq:Dynamics} to have its formal solution as a functional depending on the white Gaussian noises $\xi_n(t)$,
\begin{equation}
    \begin{split}
        x_n(t) &= x_n(0) +  \frac{1}{m}\int_0^t ds\; p_n(s) ,\\
        p_n(t) &= p_n(0) + \int_0^t ds\; \left[ -\frac{\partial H}{\partial x_n}(s) - \frac{\gamma_n}{m}p_n(s) + \xi_n(s)\right].
    \end{split}
    \label{eq:FormalSolution}
\end{equation}
Equation \eqref{eq:FormalSolution} implies that the functional derivatives are ${\delta x_n(t)}/{\delta \xi_m(t')}|_{t'\to t^-} = 0$ and
${\delta p_n(t)}/{\delta \xi_m(t')}|_{t'\to t^-} = \delta_{nm}$ ($\delta_{nm}$ is the usual Kronecker delta symbol). Thus we have $\expval{x_n(t) \xi_m(t)} = 0$ and $\expval{p_n(t) \xi_m(t)} = \delta_{nm}D_m$, which gives for the heat flow from the baths
\begin{equation}
    \expval{j^B_n} = \frac{1}{m} \left[ D_n-\gamma_n \frac{\expval{p_n^2}}{m} \right].
\end{equation}
In all simulations we check that $|J_L|=|J_R|$ within the numerical tolerance of the computer. To measure the asymmetry of the heat currents we use the rectification factor $R$ defined as
\begin{equation}
    R = \frac{ J_\rightarrow - J_\leftarrow}{max(J_\rightarrow,J_\leftarrow)}.
    \label{eq:R_Factor}
\end{equation}
$R$ values may go from $-1$ to $1$. If there is no rectification $J_\rightarrow = J_\leftarrow $ and $R=0$. For perfect rectification in the right (left) direction, $J_\rightarrow \gg J_\leftarrow$ ($J_\rightarrow \ll J_\leftarrow$), and $R = 1$ ($R = -1$).
\subsection{Algebraic, small-oscillations approach to calculate the steady state\label{steadyState}}
To find the temperature profiles and heat currents in the steady state the usual approach is to solve the SDE system in Eq. \eqref{eq:Dynamics} up to long times  and for many realizations of the white noises $\bm\xi (t)$. In that way the ensemble averages $\expval{p_n(t \to \infty)^2}$, necessary for both the temperature profiles and heat currents, are computed. This standard route implies a heavy computational effort, in particular  when we want to study the heat transport for several bath configurations, frequency gradients and chain parameters. It is possible to circumvent this difficulty and find ensemble averages like $\expval{x_n x_m}$, $\expval{x_n p_m}$, $\expval{p_n p_m}$ (second order moments) without integrating any SDE \cite{Sarkka2019}. The idea is to impose the condition $\frac{d\expval{\cdot\cdot\cdot}}{dt}$ = 0 for all the second order moments and linearize the dynamical equations of the system around equilibrium.
A system of linear algebraic equations for the moments results, that can be easily solved without solving the SDE many times.
% to solve a system of linear algebraic equations. We will analyze the validity of the linearized approximation in the following section.

To linearize the SDE in Eq. \eqref{eq:Dynamics} we approximate the potential energy of the Hamiltonian in Eq. \eqref{eq:ChainHamiltonian}, $V(\bm{x}) = V_{int}(\bm{x}) + m\;\sum_n \omega_n^2 (x_n-x_n^{(0)})^2/2$, by its harmonic approximation around the equilibrium positions $\bm{x}^{eq}$, defined by  $\frac{\partial V(\bm{x})}{\partial\bm{x}}\Big|_{\bm{x}=\bm{x}^{eq}} = 0$. The approximate potential (ignoring the zero-point energy) is
\begin{equation}
    V(\bm{x})\approx  \frac{1}{2} \sum_{n,m} K_{nm} (x_n-x_n^{eq})(x_m-x_m^{eq}),
\end{equation}
with $K_{nm} = \frac{\partial^2 V(\bm{x})}{\partial x_n \partial x_m}\Big|_{\bm{x}=\bm{x}^{eq}}$ being the Hessian matrix entries of $V(\bm{x})$ around the equilibrium configuration \cite{James1998}
\begin{equation}
    K_{nm} =
    \begin{cases}
        m \omega_n^2 + 2 \left(\frac{q^2}{4\pi\varepsilon_0}\right) \sum_{l \neq n  }\frac{1}{\left|x_n^{eq}-x_l^{eq}\right|^3} & \text{if  } n=m\\

         - 2 \left(\frac{q^2}{4\pi\varepsilon_0}\right) \frac{1}{\left|x_n^{eq}-x_m^{eq}\right|^3} & \text{if  } n \neq m
    \end{cases}.
\end{equation}
Note that this approximation does not modify the two main features of the system, namely asymmetry and long range interactions, which are manifest in the asymmetric distribution of $\omega_n$ and the non-zero off-diagonal elements of the $K$ matrix, respectively. In the following we will use $y_n=x_n-x_n^{eq}$ to simplify the notation. The linearized dynamics around the equilibrium positions are given by
\begin{equation}
    \begin{split}
        \dot{y}_n &= \frac{1}{m}p_n,\\
        \dot{p}_n &= -\sum_{l}K_{nl}y_l- \frac{\gamma_n}{m}p_n + \xi_n(t).
    \end{split}
    \label{eq:DynamicsHarmonic}
\end{equation}
Now, we set ${d\expval{\cdot\!\cdot\!\cdot}}/{dt} = 0$ for all the moments. Using Eq. \eqref{eq:DynamicsHarmonic} and applying Novikov's theorem we get
%
% \begin{widetext}
% \begin{eqnarray}
%     \frac{d}{dt}(x_n x_l) &=& \frac{1}{m}(x_n p_l + x_l p_n) , \nonumber\\
%     \frac{d}{dt}(x_n p_l) &=& \frac{1}{m}\left(p_n p_l - \gamma_l x_n p_l\right) - \sum_m K_{lm}x_n x_m + x_n \xi_l , \nonumber\\
%     \frac{d}{dt}(p_n p_l) &=& - \sum_m x_m\left( K_{nm}p_l + K_{lm} p_n \right) - \frac{1}{m}(\gamma_l + \gamma_n) p_n p_l + \xi_n p_l + \xi_l p_n .
% \label{eq:DynamicsHarmonicOfMomenta}
% \end{eqnarray}
% \end{widetext}
%
% Now we make use of $\frac{d\expval{}}{dt} = 0$ and the Novikov theorem \cite{1965JETP20.1290N,PhysRevB.83.134418}, which in this case implies $\expval{ x_n \xi_l } = 0 $ and $\expval{ p_n \xi_l } = \delta_{nl}D_n$, to finally find
%
\begin{widetext}
    \begin{equation}
        \begin{split}
            \expval{p_n p_l} - \gamma_l\expval{y_n p_l} - \sum_m K_{lm}\expval{y_n y_m} &= 0,\\
            \sum_{m}\left[ K_{nm}\expval{y_m p_l} + K_{lm}\expval{y_m p_n} \right] + \frac{1}{m}\left( \gamma_l + \gamma_n \right)\expval{p_n p_l} &= 2 \delta_{nl} D_n.
        \end{split}
        \label{eq:SteadyStateEquation}
    \end{equation}
\end{widetext}
The system \eqref{eq:SteadyStateEquation} is linear in the second order moments so it can be solved numerically to find the steady-state values of the moments. Besides Eq. \eqref{eq:SteadyStateEquation} we have that $\expval{y_n p_l} = - \expval{y_l p_n}$, which follows from Eq. \eqref{eq:DynamicsHarmonic} and $d\expval{y_n y_m}/dt = 0$. Since there are $\frac{1}{2}N(N-1)$ independent $\expval{y_n p_l}$ moments, we choose the ones with $n<l$. Similarly, the moments $\expval{y_n y_l}$ and $\expval{p_n p_l}$ contribute with $\frac{1}{2}N(N+1)$ independent variables each and we choose the ones with $n\leq m$. Thus there are in total $\frac{1}{2}N(3N+1)$ independent moments that we arrange in the vector
\begin{equation}
\begin{split}
    \bm\eta = \Big[ &\expval{y_1 y_1},\expval{y_1 y_2},...,\expval{y_N y_N},\\
    &\expval{p_1 p_1},\expval{p_1 p_2},...,\expval{p_N p_N},\\
    &\expval{y_1 p_2},\expval{y_1 p_3},...,\expval{y_{N-1} p_N}\Big]^T
    \label{eq:CovariancesVector}
\end{split}
\end{equation}
There are the same number of independent equations
%in Eq. \eqref{eq:SteadyStateEquation}
as independent moments: $N^2$
equations correspond to the first line in Eq.  \eqref{eq:SteadyStateEquation}, and $\frac{1}{2}N(N+1)$ equations
to the second line because of the symmetry with respect to $n,l$. The system of equations \eqref{eq:SteadyStateEquation} may be compactly written as $\mathbf{A}\boldsymbol\eta = \mathbf{B}$, where $\mathbf{A}$ and $\mathbf{B}$ are a $\frac{1}{2}N(3N+1)$ square matrix and vector.
% which are built by rearranging the coefficients in the left and right hand sizes of \eqref{eq:SteadyStateEquation} in the appropiate way.
%
%
%
%
%
\section{Numerical Results\label{Numerical Results}}
\begin{figure}[h]
    \includegraphics[width=\linewidth]{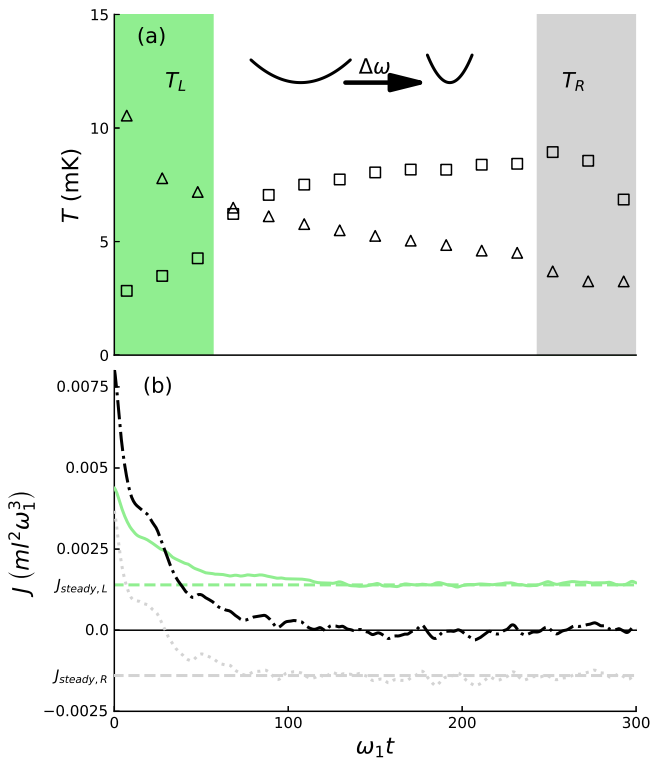}
 \caption{(Color online)(a) Temperatures of the ions in the stationary state for a graded chain with the parameters described in section \ref{Results_A}. The temperature profiles found with the algebraic Method (Eq. \eqref{eq:SteadyStateEquation}) are indistinguishable from the ones found solving the Langevin equation (Eq. \eqref{eq:Dynamics}). Empty triangles (squares) correspond to $T_L = T_H$ ($T_L = T_C$) and $T_R = T_C$ ($T_R = T_H$). (b) Heat current along the chain as a function of time for $T_L = T_H$ and $T_R = T_C$. The Solid green line corresponds to the heat current coming from the left reservoir into the chain, the dotted grey line corresponds to the heat current from the chain into the right reservoir. In the steady state, the sum of the incoming and outgoing heat currents, dotted-dashed black line, must cancel. Dashed horizontal green (grey) line corresponds to the heat current from the left reservoir (into the right reservoir) predicted by the steady state equation.}
    \label{fig:Temperature_Profiles_Magnesium}
\end{figure}

In this section we display the results of our simulations. To find the temperature profiles and the currents in the steady state we use the algebraic method described in section \ref{steadyState}. We also check that the results coincide with solving Eq. \eqref{eq:Dynamics} for many different realizations of the noisy forces $\bm\xi (t)$ and averaging. The code for all the numerical simulations have been written in the language \textit{Julia} \cite{Bezanson2012,Bezanson2017}. In particular, to solve the Langevin equation, we used \textit{Julia}'s package \textit{DifferentialEquations.jl} \cite{Rackauckas2017}.

To model the baths and the chain we use atomic data taken from ion trap experiments \cite{Leupold2015,Lo2015}. We consider 15 $^{24}$Mg$^+$ ions. The three leftmost and three rightmost ions are illuminated by Doppler cooling lasers. The Doppler cooling lasers excite the transition $3s^2S_{1/2}\rightarrow 3p^2P_{1/2}$, with angular frequency $\omega_0 = 2 \pi \times 1069$ THz and excited state line width $\Gamma = 2\pi \times 41.3$ MHz \cite{Ruiz2014}. For this ionic species and atomic transition the Doppler limit is $T_D = 1$ mK. %We use typical trap frequencies of the order of MegaHertzs and intertrap spacings of a few tens of $\mu$m.
The intensities of the laser beams are small compared to the saturation intensity $I_0$ so that Eq. \eqref{eq:DopplerCooling} holds. We take $I_n/I_0 = 0.08$ for the ions  in the laser beams, whereas  $I_n=0$ for the rest.

 The temperatures $T_L,\,T_R$ of the left and right laser baths are controlled with their detunings $\delta_L,\,\delta_R$ with respect to the atomic transition. We fix two values for the detunings, $\delta_H$ and $\delta_C$, such that $T_H>T_C$ (hot and cold baths, also source and drain) and we define $J_\rightarrow$ ($J_\leftarrow$) as the heat current in the chain when $T_L = T_H$ and $T_R = T_C$ ($T_L = T_C$ and $T_R = T_H$).

In this section we consider two frequency profiles for the traps: graded or segmented. In the graded chain the frequency increases by $\frac{\Delta\omega}{N-1}$ from one trap to the next. If the frequency of the leftmost trap is $\omega_1$, the frequency of the $n$th trap will be $\omega_n = \omega_1 +\Delta\omega\frac{n-1}{N-1}$. The frequency of the rightmost trap is $\omega_1 +\Delta\omega$. In the segmented chain, the left half of the chain has trapping frequencies $\omega_1$ while the other half has $\omega_1 +\Delta\omega$.

\subsection{Evolution to steady state \label{Results_A}}

\begin{figure}[h]
    \includegraphics[width=\linewidth]{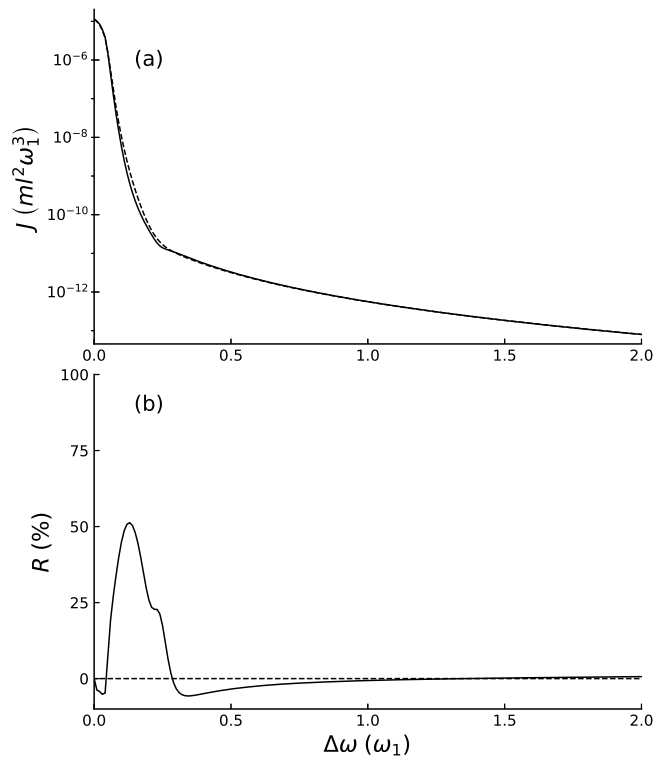}
    \caption{Graded chain of $N=15$ $^{24}$Mg$^+$ ions. (a) Fluxes for different frequency gradients. The dashed line corresponds to the right-going flux $J_\rightarrow$, i.e., when $T_L = T_H$ and $T_R = T_C$.; the solid line corresponds to the left-going flux $J_\leftarrow$, i.e., when $T_L = T_C$ and $T_R = T_H$. (b) rectification factor. Parameters in this figure are $\omega_1 = 2 \pi \times 1$ MHz, $l = 5.25\;\mu$m, $a = 4.76\, l$ ($25\,\mu$m), $\delta_H = -0.02 \,\Gamma$, and $\delta_C = -0.1 \, \Gamma$.}
    \label{fig:Graded_24Mg_FluxAndRectification_VS_FreqGradient}
\end{figure}

To compare the results by solving Eq. \eqref{eq:Dynamics} and averaging and those from the algebraic method we  simulated a frequency graded chain with a lattice constant (intertrap distance) $a = 50\,\mu$m and a trapping frequency $\omega_1 = 2\pi \times 50$ kHz for the leftmost ion, Fig. \ref{fig:Temperature_Profiles_Magnesium}.  The number of ions interacting with the laser beams (three on each bath) is consistent with the lattice constant and typical waists of Gaussian laser beams \cite{Leupold2015,Lo2015}. To set the trap distance we fix first the characteristic length  $l =  \left(\frac{q^2}{4\pi\varepsilon_0}\frac{1}{m\omega_1^2}\right)^{1/3}$ as the distance for which the Coulomb repulsion of two ions equals the trap  potential energy for an ion at a distance
$l$ away from the center of its trap.
If $a<l$, the Coulomb repulsion of the ions is stronger than the trap confinement which makes the ions jump from their traps. With the parameters used in this section we have $l = 38.7\,\mu$m ($a = 1.29 \,l$). The detunings of the \textit{hot} and \textit{cold} lasers are $\delta_H = -0.02 \, \Gamma $ and $\delta_C = -0.1 \, \Gamma$ which gives temperatures $T_H \approx 12$ mK and $T_C \approx 3$ mK. We fix the value $\Delta\omega = 0.5 \, \omega_1$ for the frequency gradient.

The results of the two methods are in very good agreement. In the scale of Fig. \ref{fig:Temperature_Profiles_Magnesium} (a)
the calculated local temperatures are undistinguishable. In the calculation based on solving the dynamics we had to integrate \eqref{eq:Dynamics} for $N_{trials} = 1000$ realizations of white noise $\bm{\xi}(t)$. The method based on the system of moments
shortened the calculation time with respect to the dynamical trajectories   by a factor of $1/700$. In fact, the time gain is even more important because
the dynamical method requires further processing, performing a time averaging to compute the stationary flux in addition to noise averaging, see Fig.  \ref{fig:Temperature_Profiles_Magnesium} (b).

Additionally, the relaxation to the steady state slows down when the frequencies of the traps increase since the deterministic part of the Langevin equation dominates the dynamics over the stochastic part, entering an under-damped regime. In contrast, this increase does not affect the
algebraic method.
%%%%%%%%%%%%%%%%%%%%
\begin{figure}[t]
    \includegraphics[width=\linewidth]{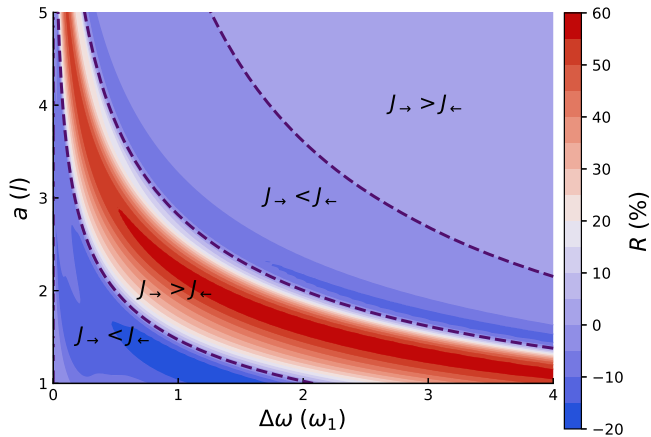}
    \caption{(Color online) Rectification factor in a graded chain of $N=15$ $^{24}$Mg$^+$ ions for different trap distances and frequency gradient. The dashed lines are for $R = 0$ and delimit the regions $J_\rightarrow > J_\leftarrow$ and $J_\rightarrow < J_\leftarrow$. The parameters  are $\omega_1 = 2 \pi \times 1$ MHz, $l = 5.25\,\mu$m, $\delta_H = -0.02 \,\Gamma$, and $\delta_C = -0.1 \, \Gamma$.}
    \label{fig:Graded_24Mg_Rectification_VS_Gradient_and_lattConstant}
\end{figure}
\subsection{Rectification in frequency graded chains \label{GradedChains}}
In this section we display the results of some simulations that demonstrate rectification. We  used the method described in section \ref{steadyState} for $^{24}$Mg$^+$ ions with the same parameters for the baths used before. We fix the trapping frequency of the leftmost trap to $\omega_1 = 2\pi \times 1$ MHz, which sets a characteristic length $l = 5.25\,\mu$m, and a trap spacing $a = 4.76\, l$ ($25\,\mu$m). Figure \ref{fig:Graded_24Mg_FluxAndRectification_VS_FreqGradient} shows the results with these parameters in a graded chain. Figure \ref{fig:Graded_24Mg_FluxAndRectification_VS_FreqGradient} (a) shows that both the right-going flux (dashed line) and the left-going flux (solid line) decrease rapidly as the frequency gradient is increased.
%In Fig. \ref{fig:Graded_24Mg_FluxAndRectification_VS_FreqGradient} we see that
The rectification reaches its maximum value for a frequency gradient of $\Delta\omega \approx 0.1 \omega_1$. The fluxes cross so there are some points where the rectification is exactly zero, besides the trivial one at $\Delta\omega = 0$, at $\Delta\omega = 0.05\,\omega_1,\;0.3\,\omega_1,\;1.3\,\omega_1$. At these points the direction of rectification reverses, presumably as a consequence of the changes in the match/mismatch of the temperature dependent local power spectra.
The change of rectification direction occurs for all the choices of parameters, as displayed in Fig. \ref{fig:Graded_24Mg_Rectification_VS_Gradient_and_lattConstant}. Figure \ref{fig:Graded_24Mg_Rectification_VS_Gradient_and_lattConstant} gives the rectification factor for different trap distances and frequency gradients.  $0$-rectification curves separate regions with different rectification direction. The second region in Fig. \ref{fig:Graded_24Mg_Rectification_VS_Gradient_and_lattConstant} (starting from the left) would be the most interesting one to build a thermal diode, since rectification reaches its largest values there.
%In the other regions, rectification values are small, around $10\%$, and thus not so useful for applications.

\begin{figure}[t]
    \includegraphics[width=\linewidth]{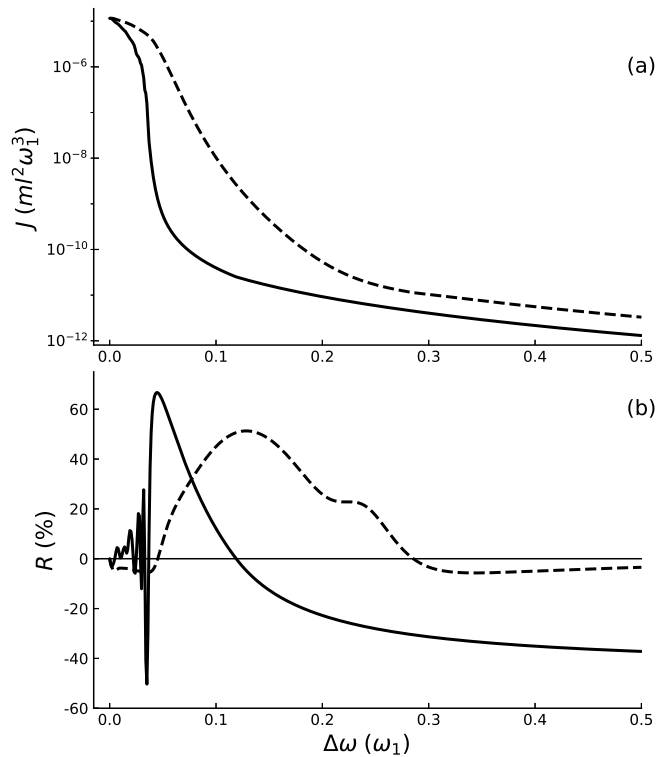}
    \caption{Comparison of graded and segmented frequency profiles in a chain of $N=15$ $^{24}$Mg$^+$ ions. (a) Maximum
    of $J_\rightarrow$ and $J_\leftarrow$
     for the graded and segmented chain for different frequency gradients. (b) Rectification factor. Dashed lines: graded chain; solid lines: segmented chain. The parameters are $\omega_1 = 2 \pi \times 1$ MHz, $l = 5.25\,\mu$m, $a = 4.76 \, l$, $\delta_H = -0.02 \, \Gamma$, and $\delta_C = -0.1 \, \Gamma$.}
    \label{fig:Graded_VS_Segmented}
\end{figure}

We have also compared the performance of the graded thermal diode against a segmented version of it.
%In the segmented chain, the traps on the left have a trapping frequency $\omega_1$ and the ions on the right $\omega_1 + \Delta\omega$.
%, as explained in the second paragraph at the beginning of section \ref{Numerical Results}.
Even though the optimal rectification  in Fig. \ref{fig:Graded_VS_Segmented} (a)  for the segmented chain is larger
than for the graded chain, the fluxes are generally much larger for the graded chain, see Fig. \ref{fig:Graded_VS_Segmented} (b),
which makes it more interesting for
applications.
%the key point is that the fluxes themselves are much smaller compares the results for graded and segmented diodes. Figure \ref{fig:Graded_VS_Segmented}(a) displays a faster decrease of the heat flux  with $\Delta\omega$ for the segmented diode. We see in Fig. \ref{fig:Graded_VS_Segmented} (b) that the segmented chain can produce a larger maximum rectification factor, but the graded one is more robust against perturbations of $\Delta\omega$ from its optimal value. For larger values of the frequency gradient the rectification of the graded chain goes to zero while for the segmented one it stabilizes to a non-zero value, however, it is not an useful feature since, for such large gradients the heat flux decreases several orders of magnitude.
%
%
%
%
\section{Conclusions \label{Conclusions}}
In this article we have numerically demonstrated heat rectification in a chain of ions trapped in individual microtraps with graded frequencies, connected at both ends to thermal baths
created by optical molasses. An alternative to implement a graded
frequency profile in the lab
could be combining a collective Paul trap for all the ions with on-site dipolar laser forces \cite{Freitas2016,Enderlein2012,Bermudez2013,Schneider2010}.

A goal of this article is to connect two communities, ion trappers and
researchers on heat-rectification models. The results found are encouraging and demonstrate the potential of a trapped-ion platform to experimentally investigate heat rectification schemes. Trapped ions are interesting to this end because they are highly controllable, and may easily adopt several features to enhance rectification, such as
the ones explored here (long-range interactions and an asymmetrical gradation),
or others such as time dependent forces \cite{Li2012,Riera-Campeny2018}, or different nonlinearities in onsite forces.

The calculation of the steady state has been performed with an algebraic approach much faster than
the time-consuming integration and averaging over noise and time of the dynamical equations.
The algebraic approach linearizes the forces around equilibrium positions which, in this system and for the realistic parameters considered  is well justified and tested numerically.
The results found provide additional evidence that simple linear models may  rectify heat flow \cite{Pereira2017}.  We  underline that our linear
model is, arguably,  even simpler than some linear ``minimalist, toy models'' in \cite{Pereira2017} that showed rectification (our on-site forces are already linear from the start and the temperature dependence is only in the coefficients of the Langevin baths), with the important bonus of being also realistic.

\begin{acknowledgments}
The authors would like to thank Joseba Alonso and Daniel  Alonso for fruitful discussions and comments. M. A. Sim\'on acknowledges support by the Basque Government predoctoral program (Grant no. PRE-2018-2-0177). We thank funding from Grant  No. PGC2018-101355-B-I00 and Fis2016-80681-P (MCIU/AEI/FEDER,UE),
and Basque Government (Grant No. IT986- 16).

\end{acknowledgments}

\bibliographystyle{apsrev4-1.bst}
\bibliography{Bibliography.bib}

%merlin.mbs apsrev4-1.bst 2010-07-25 4.21a (PWD, AO, DPC) hacked
%Control: key (0)
%Control: author (72) initials jnrlst
%Control: editor formatted (1) identically to author
%Control: production of article title (-1) disabled
%Control: page (0) single
%Control: year (1) truncated
%Control: production of eprint (0) enabled
\begin{thebibliography}{58}%
\makeatletter
\providecommand \@ifxundefined [1]{%
 \@ifx{#1\undefined}
}%
\providecommand \@ifnum [1]{%
 \ifnum #1\expandafter \@firstoftwo
 \else \expandafter \@secondoftwo
 \fi
}%
\providecommand \@ifx [1]{%
 \ifx #1\expandafter \@firstoftwo
 \else \expandafter \@secondoftwo
 \fi
}%
\providecommand \natexlab [1]{#1}%
\providecommand \enquote  [1]{``#1''}%
\providecommand \bibnamefont  [1]{#1}%
\providecommand \bibfnamefont [1]{#1}%
\providecommand \citenamefont [1]{#1}%
\providecommand \href@noop [0]{\@secondoftwo}%
\providecommand \href [0]{\begingroup \@sanitize@url \@href}%
\providecommand \@href[1]{\@@startlink{#1}\@@href}%
\providecommand \@@href[1]{\endgroup#1\@@endlink}%
\providecommand \@sanitize@url [0]{\catcode `\\12\catcode `\$12\catcode
  `\&12\catcode `\#12\catcode `\^12\catcode `\_12\catcode `\%12\relax}%
\providecommand \@@startlink[1]{}%
\providecommand \@@endlink[0]{}%
\providecommand \url  [0]{\begingroup\@sanitize@url \@url }%
\providecommand \@url [1]{\endgroup\@href {#1}{\urlprefix }}%
\providecommand \urlprefix  [0]{URL }%
\providecommand \Eprint [0]{\href }%
\providecommand \doibase [0]{http://dx.doi.org/}%
\providecommand \selectlanguage [0]{\@gobble}%
\providecommand \bibinfo  [0]{\@secondoftwo}%
\providecommand \bibfield  [0]{\@secondoftwo}%
\providecommand \translation [1]{[#1]}%
\providecommand \BibitemOpen [0]{}%
\providecommand \bibitemStop [0]{}%
\providecommand \bibitemNoStop [0]{.\EOS\space}%
\providecommand \EOS [0]{\spacefactor3000\relax}%
\providecommand \BibitemShut  [1]{\csname bibitem#1\endcsname}%
\let\auto@bib@innerbib\@empty
%</preamble>
\bibitem [{\citenamefont {Starr}(1936)}]{Starr1936}%
  \BibitemOpen
  \bibfield  {author} {\bibinfo {author} {\bibfnamefont {C.}~\bibnamefont
  {Starr}},\ }\href@noop {} {\bibfield  {journal} {\bibinfo  {journal}
  {Physics}\ }\textbf {\bibinfo {volume} {7}},\ \bibinfo {pages} {15} (\bibinfo
  {year} {1936})}\BibitemShut {NoStop}%
\bibitem [{\citenamefont {Terraneo}\ \emph {et~al.}(2002)\citenamefont
  {Terraneo}, \citenamefont {Peyrard},\ and\ \citenamefont
  {Casati}}]{Terraneo2002}%
  \BibitemOpen
  \bibfield  {author} {\bibinfo {author} {\bibfnamefont {M.}~\bibnamefont
  {Terraneo}}, \bibinfo {author} {\bibfnamefont {M.}~\bibnamefont {Peyrard}}, \
  and\ \bibinfo {author} {\bibfnamefont {G.}~\bibnamefont {Casati}},\ }\href
  {\doibase 10.1103/PhysRevLett.88.094302} {\bibfield  {journal} {\bibinfo
  {journal} {Phys. Rev. Lett.}\ }\textbf {\bibinfo {volume} {88}},\ \bibinfo
  {pages} {094302} (\bibinfo {year} {2002})}\BibitemShut {NoStop}%
\bibitem [{\citenamefont {Pereira}(2019)}]{Pereira2019}%
  \BibitemOpen
  \bibfield  {author} {\bibinfo {author} {\bibfnamefont {E.}~\bibnamefont
  {Pereira}},\ }\href {\doibase 10.1209/0295-5075/126/14001} {\bibfield
  {journal} {\bibinfo  {journal} {{EPL} (Europhysics Letters)}\ }\textbf
  {\bibinfo {volume} {126}},\ \bibinfo {pages} {14001} (\bibinfo {year}
  {2019})}\BibitemShut {NoStop}%
\bibitem [{\citenamefont {Roberts}\ and\ \citenamefont
  {Walker}(2011)}]{Roberts2011}%
  \BibitemOpen
  \bibfield  {author} {\bibinfo {author} {\bibfnamefont {N.}~\bibnamefont
  {Roberts}}\ and\ \bibinfo {author} {\bibfnamefont {D.}~\bibnamefont
  {Walker}},\ }\href {\doibase
  https://doi.org/10.1016/j.ijthermalsci.2010.12.004} {\bibfield  {journal}
  {\bibinfo  {journal} {International Journal of Thermal Sciences}\ }\textbf
  {\bibinfo {volume} {50}},\ \bibinfo {pages} {648 } (\bibinfo {year}
  {2011})}\BibitemShut {NoStop}%
\bibitem [{\citenamefont {Li}\ \emph {et~al.}(2012)\citenamefont {Li},
  \citenamefont {Ren}, \citenamefont {Wang}, \citenamefont {Zhang},
  \citenamefont {H\"anggi},\ and\ \citenamefont {Li}}]{Li2012}%
  \BibitemOpen
  \bibfield  {author} {\bibinfo {author} {\bibfnamefont {N.}~\bibnamefont
  {Li}}, \bibinfo {author} {\bibfnamefont {J.}~\bibnamefont {Ren}}, \bibinfo
  {author} {\bibfnamefont {L.}~\bibnamefont {Wang}}, \bibinfo {author}
  {\bibfnamefont {G.}~\bibnamefont {Zhang}}, \bibinfo {author} {\bibfnamefont
  {P.}~\bibnamefont {H\"anggi}}, \ and\ \bibinfo {author} {\bibfnamefont
  {B.}~\bibnamefont {Li}},\ }\href {\doibase 10.1103/RevModPhys.84.1045}
  {\bibfield  {journal} {\bibinfo  {journal} {Rev. Mod. Phys.}\ }\textbf
  {\bibinfo {volume} {84}},\ \bibinfo {pages} {1045} (\bibinfo {year}
  {2012})}\BibitemShut {NoStop}%
\bibitem [{\citenamefont {Ye}\ and\ \citenamefont {Cao}(2017)}]{Ye2017}%
  \BibitemOpen
  \bibfield  {author} {\bibinfo {author} {\bibfnamefont {Z.-Q.}\ \bibnamefont
  {Ye}}\ and\ \bibinfo {author} {\bibfnamefont {B.-Y.}\ \bibnamefont {Cao}},\
  }\href {\doibase 10.1039/C7NR02696J} {\bibfield  {journal} {\bibinfo
  {journal} {Nanoscale}\ }\textbf {\bibinfo {volume} {9}},\ \bibinfo {pages}
  {11480} (\bibinfo {year} {2017})}\BibitemShut {NoStop}%
\bibitem [{\citenamefont {Wang}\ and\ \citenamefont {Li}(2008)}]{Wang2008}%
  \BibitemOpen
  \bibfield  {author} {\bibinfo {author} {\bibfnamefont {L.}~\bibnamefont
  {Wang}}\ and\ \bibinfo {author} {\bibfnamefont {B.}~\bibnamefont {Li}},\
  }\href {\doibase 10.1103/PhysRevLett.101.267203} {\bibfield  {journal}
  {\bibinfo  {journal} {Phys. Rev. Lett.}\ }\textbf {\bibinfo {volume} {101}},\
  \bibinfo {pages} {267203} (\bibinfo {year} {2008})}\BibitemShut {NoStop}%
\bibitem [{\citenamefont {Wang}\ and\ \citenamefont {Li}(2007)}]{Wang2007}%
  \BibitemOpen
  \bibfield  {author} {\bibinfo {author} {\bibfnamefont {L.}~\bibnamefont
  {Wang}}\ and\ \bibinfo {author} {\bibfnamefont {B.}~\bibnamefont {Li}},\
  }\href {\doibase 10.1103/PhysRevLett.99.177208} {\bibfield  {journal}
  {\bibinfo  {journal} {Phys. Rev. Lett.}\ }\textbf {\bibinfo {volume} {99}},\
  \bibinfo {pages} {177208} (\bibinfo {year} {2007})}\BibitemShut {NoStop}%
\bibitem [{\citenamefont {Li}\ \emph {et~al.}(2006)\citenamefont {Li},
  \citenamefont {Wang},\ and\ \citenamefont {Casati}}]{Casati2006}%
  \BibitemOpen
  \bibfield  {author} {\bibinfo {author} {\bibfnamefont {B.}~\bibnamefont
  {Li}}, \bibinfo {author} {\bibfnamefont {L.}~\bibnamefont {Wang}}, \ and\
  \bibinfo {author} {\bibfnamefont {G.}~\bibnamefont {Casati}},\ }\href
  {\doibase 10.1063/1.2191730} {\bibfield  {journal} {\bibinfo  {journal}
  {Applied Physics Letters}\ }\textbf {\bibinfo {volume} {88}},\ \bibinfo
  {pages} {143501} (\bibinfo {year} {2006})}\BibitemShut {NoStop}%
\bibitem [{\citenamefont {Joulain}\ \emph {et~al.}(2016)\citenamefont
  {Joulain}, \citenamefont {Drevillon}, \citenamefont {Ezzahri},\ and\
  \citenamefont {Ordonez-Miranda}}]{Joulain2016}%
  \BibitemOpen
  \bibfield  {author} {\bibinfo {author} {\bibfnamefont {K.}~\bibnamefont
  {Joulain}}, \bibinfo {author} {\bibfnamefont {J.}~\bibnamefont {Drevillon}},
  \bibinfo {author} {\bibfnamefont {Y.}~\bibnamefont {Ezzahri}}, \ and\
  \bibinfo {author} {\bibfnamefont {J.}~\bibnamefont {Ordonez-Miranda}},\
  }\href {\doibase 10.1103/PhysRevLett.116.200601} {\bibfield  {journal}
  {\bibinfo  {journal} {Phys. Rev. Lett.}\ }\textbf {\bibinfo {volume} {116}},\
  \bibinfo {pages} {200601} (\bibinfo {year} {2016})}\BibitemShut {NoStop}%
\bibitem [{\citenamefont {Chang}\ \emph {et~al.}(2006)\citenamefont {Chang},
  \citenamefont {Okawa}, \citenamefont {Majumdar},\ and\ \citenamefont
  {Zettl}}]{Chang2006}%
  \BibitemOpen
  \bibfield  {author} {\bibinfo {author} {\bibfnamefont {C.~W.}\ \bibnamefont
  {Chang}}, \bibinfo {author} {\bibfnamefont {D.}~\bibnamefont {Okawa}},
  \bibinfo {author} {\bibfnamefont {A.}~\bibnamefont {Majumdar}}, \ and\
  \bibinfo {author} {\bibfnamefont {A.}~\bibnamefont {Zettl}},\ }\href
  {\doibase 10.1126/science.1132898} {\bibfield  {journal} {\bibinfo  {journal}
  {Science}\ }\textbf {\bibinfo {volume} {314}},\ \bibinfo {pages} {1121}
  (\bibinfo {year} {2006})}\BibitemShut {NoStop}%
\bibitem [{\citenamefont {Li}\ \emph {et~al.}(2008)\citenamefont {Li},
  \citenamefont {H\"{a}nggi},\ and\ \citenamefont {Li}}]{Li2008}%
  \BibitemOpen
  \bibfield  {author} {\bibinfo {author} {\bibfnamefont {N.}~\bibnamefont
  {Li}}, \bibinfo {author} {\bibfnamefont {P.}~\bibnamefont {H\"{a}nggi}}, \
  and\ \bibinfo {author} {\bibfnamefont {B.}~\bibnamefont {Li}},\ }\href
  {http://stacks.iop.org/0295-5075/84/i=4/a=40009} {\bibfield  {journal}
  {\bibinfo  {journal} {EPL (Europhysics Letters)}\ }\textbf {\bibinfo {volume}
  {84}},\ \bibinfo {pages} {40009} (\bibinfo {year} {2008})}\BibitemShut
  {NoStop}%
\bibitem [{\citenamefont {Hu}\ \emph {et~al.}(2006)\citenamefont {Hu},
  \citenamefont {Yang},\ and\ \citenamefont {Zhang}}]{Hu2006}%
  \BibitemOpen
  \bibfield  {author} {\bibinfo {author} {\bibfnamefont {B.}~\bibnamefont
  {Hu}}, \bibinfo {author} {\bibfnamefont {L.}~\bibnamefont {Yang}}, \ and\
  \bibinfo {author} {\bibfnamefont {Y.}~\bibnamefont {Zhang}},\ }\href
  {\doibase 10.1103/PhysRevLett.97.124302} {\bibfield  {journal} {\bibinfo
  {journal} {Phys. Rev. Lett.}\ }\textbf {\bibinfo {volume} {97}},\ \bibinfo
  {pages} {124302} (\bibinfo {year} {2006})}\BibitemShut {NoStop}%
\bibitem [{\citenamefont {Zeng}\ and\ \citenamefont {Wang}(2008)}]{Zeng2008}%
  \BibitemOpen
  \bibfield  {author} {\bibinfo {author} {\bibfnamefont {N.}~\bibnamefont
  {Zeng}}\ and\ \bibinfo {author} {\bibfnamefont {J.-S.}\ \bibnamefont
  {Wang}},\ }\href {\doibase 10.1103/PhysRevB.78.024305} {\bibfield  {journal}
  {\bibinfo  {journal} {Phys. Rev. B}\ }\textbf {\bibinfo {volume} {78}},\
  \bibinfo {pages} {024305} (\bibinfo {year} {2008})}\BibitemShut {NoStop}%
\bibitem [{\citenamefont {Katz}\ and\ \citenamefont
  {Kosloff}(2016)}]{Katz2016}%
  \BibitemOpen
  \bibfield  {author} {\bibinfo {author} {\bibfnamefont {G.}~\bibnamefont
  {Katz}}\ and\ \bibinfo {author} {\bibfnamefont {R.}~\bibnamefont {Kosloff}},\
  }\href {\doibase 10.3390/e18050186} {\bibfield  {journal} {\bibinfo
  {journal} {Entropy}\ }\textbf {\bibinfo {volume} {18}} (\bibinfo {year}
  {2016}),\ 10.3390/e18050186}\BibitemShut {NoStop}%
\bibitem [{\citenamefont {Benenti}\ \emph {et~al.}(2016)\citenamefont
  {Benenti}, \citenamefont {Casati}, \citenamefont {Mej{\'\i}a-Monasterio},\
  and\ \citenamefont {Peyrard}}]{Benenti2016}%
  \BibitemOpen
  \bibfield  {author} {\bibinfo {author} {\bibfnamefont {G.}~\bibnamefont
  {Benenti}}, \bibinfo {author} {\bibfnamefont {G.}~\bibnamefont {Casati}},
  \bibinfo {author} {\bibfnamefont {C.}~\bibnamefont {Mej{\'\i}a-Monasterio}},
  \ and\ \bibinfo {author} {\bibfnamefont {M.}~\bibnamefont {Peyrard}},\
  }\enquote {\bibinfo {title} {From thermal rectifiers to thermoelectric
  devices},}\ in\ \href {\doibase 10.1007/978-3-319-29261-8_10} {\emph
  {\bibinfo {booktitle} {Thermal Transport in Low Dimensions: From Statistical
  Physics to Nanoscale Heat Transfer}}},\ \bibinfo {editor} {edited by\
  \bibinfo {editor} {\bibfnamefont {S.}~\bibnamefont {Lepri}}}\ (\bibinfo
  {publisher} {Springer International Publishing},\ \bibinfo {address} {Cham},\
  \bibinfo {year} {2016})\ pp.\ \bibinfo {pages} {365--407}\BibitemShut
  {NoStop}%
\bibitem [{\citenamefont {Li}\ \emph {et~al.}(2004)\citenamefont {Li},
  \citenamefont {Wang},\ and\ \citenamefont {Casati}}]{Li2004}%
  \BibitemOpen
  \bibfield  {author} {\bibinfo {author} {\bibfnamefont {B.}~\bibnamefont
  {Li}}, \bibinfo {author} {\bibfnamefont {L.}~\bibnamefont {Wang}}, \ and\
  \bibinfo {author} {\bibfnamefont {G.}~\bibnamefont {Casati}},\ }\href
  {\doibase 10.1103/PhysRevLett.93.184301} {\bibfield  {journal} {\bibinfo
  {journal} {Phys. Rev. Lett.}\ }\textbf {\bibinfo {volume} {93}},\ \bibinfo
  {pages} {184301} (\bibinfo {year} {2004})}\BibitemShut {NoStop}%
\bibitem [{\citenamefont {Pereira}(2017)}]{Pereira2017}%
  \BibitemOpen
  \bibfield  {author} {\bibinfo {author} {\bibfnamefont {E.}~\bibnamefont
  {Pereira}},\ }\href {\doibase 10.1103/PhysRevE.96.012114} {\bibfield
  {journal} {\bibinfo  {journal} {Phys. Rev. E}\ }\textbf {\bibinfo {volume}
  {96}},\ \bibinfo {pages} {012114} (\bibinfo {year} {2017})}\BibitemShut
  {NoStop}%
\bibitem [{\citenamefont {Pons}\ \emph {et~al.}(2017)\citenamefont {Pons},
  \citenamefont {Cui}, \citenamefont {Ruschhaupt}, \citenamefont
  {Sim{\'{o}}n},\ and\ \citenamefont {Muga}}]{Pons2017}%
  \BibitemOpen
  \bibfield  {author} {\bibinfo {author} {\bibfnamefont {M.}~\bibnamefont
  {Pons}}, \bibinfo {author} {\bibfnamefont {Y.~Y.}\ \bibnamefont {Cui}},
  \bibinfo {author} {\bibfnamefont {A.}~\bibnamefont {Ruschhaupt}}, \bibinfo
  {author} {\bibfnamefont {M.~A.}\ \bibnamefont {Sim{\'{o}}n}}, \ and\ \bibinfo
  {author} {\bibfnamefont {J.~G.}\ \bibnamefont {Muga}},\ }\href {\doibase
  10.1209/0295-5075/119/64001} {\bibfield  {journal} {\bibinfo  {journal}
  {{EPL} (Europhysics Letters)}\ }\textbf {\bibinfo {volume} {119}},\ \bibinfo
  {pages} {64001} (\bibinfo {year} {2017})}\BibitemShut {NoStop}%
\bibitem [{\citenamefont {Wang}\ \emph {et~al.}(2012)\citenamefont {Wang},
  \citenamefont {Pereira},\ and\ \citenamefont {Casati}}]{Wang2012}%
  \BibitemOpen
  \bibfield  {author} {\bibinfo {author} {\bibfnamefont {J.}~\bibnamefont
  {Wang}}, \bibinfo {author} {\bibfnamefont {E.}~\bibnamefont {Pereira}}, \
  and\ \bibinfo {author} {\bibfnamefont {G.}~\bibnamefont {Casati}},\ }\href
  {\doibase 10.1103/PhysRevE.86.010101} {\bibfield  {journal} {\bibinfo
  {journal} {Phys. Rev. E}\ }\textbf {\bibinfo {volume} {86}},\ \bibinfo
  {pages} {010101} (\bibinfo {year} {2012})}\BibitemShut {NoStop}%
\bibitem [{\citenamefont {Chen}\ \emph {et~al.}(2015)\citenamefont {Chen},
  \citenamefont {Pereira},\ and\ \citenamefont {Casati}}]{Chen2015}%
  \BibitemOpen
  \bibfield  {author} {\bibinfo {author} {\bibfnamefont {S.}~\bibnamefont
  {Chen}}, \bibinfo {author} {\bibfnamefont {E.}~\bibnamefont {Pereira}}, \
  and\ \bibinfo {author} {\bibfnamefont {G.}~\bibnamefont {Casati}},\ }\href
  {http://stacks.iop.org/0295-5075/111/i=3/a=30004} {\bibfield  {journal}
  {\bibinfo  {journal} {EPL (Europhysics Letters)}\ }\textbf {\bibinfo {volume}
  {111}},\ \bibinfo {pages} {30004} (\bibinfo {year} {2015})}\BibitemShut
  {NoStop}%
\bibitem [{\citenamefont {Romero-Bastida}\ \emph {et~al.}(2017)\citenamefont
  {Romero-Bastida}, \citenamefont {Miranda-Pe\~na},\ and\ \citenamefont
  {L\'opez}}]{Romero-Bastida2017}%
  \BibitemOpen
  \bibfield  {author} {\bibinfo {author} {\bibfnamefont {M.}~\bibnamefont
  {Romero-Bastida}}, \bibinfo {author} {\bibfnamefont {J.-O.}\ \bibnamefont
  {Miranda-Pe\~na}}, \ and\ \bibinfo {author} {\bibfnamefont {J.~M.}\
  \bibnamefont {L\'opez}},\ }\href {\doibase 10.1103/PhysRevE.95.032146}
  {\bibfield  {journal} {\bibinfo  {journal} {Phys. Rev. E}\ }\textbf {\bibinfo
  {volume} {95}},\ \bibinfo {pages} {032146} (\bibinfo {year}
  {2017})}\BibitemShut {NoStop}%
\bibitem [{\citenamefont {Yang}\ \emph {et~al.}(2007)\citenamefont {Yang},
  \citenamefont {Li}, \citenamefont {Wang},\ and\ \citenamefont
  {Li}}]{Yang2007}%
  \BibitemOpen
  \bibfield  {author} {\bibinfo {author} {\bibfnamefont {N.}~\bibnamefont
  {Yang}}, \bibinfo {author} {\bibfnamefont {N.}~\bibnamefont {Li}}, \bibinfo
  {author} {\bibfnamefont {L.}~\bibnamefont {Wang}}, \ and\ \bibinfo {author}
  {\bibfnamefont {B.}~\bibnamefont {Li}},\ }\href {\doibase
  10.1103/PhysRevB.76.020301} {\bibfield  {journal} {\bibinfo  {journal} {Phys.
  Rev. B}\ }\textbf {\bibinfo {volume} {76}},\ \bibinfo {pages} {020301}
  (\bibinfo {year} {2007})}\BibitemShut {NoStop}%
\bibitem [{\citenamefont {Romero-Bastida}\ and\ \citenamefont
  {Arizmendi-Carvajal}(2013)}]{Romero-Bastida2013}%
  \BibitemOpen
  \bibfield  {author} {\bibinfo {author} {\bibfnamefont {M.}~\bibnamefont
  {Romero-Bastida}}\ and\ \bibinfo {author} {\bibfnamefont {J.~M.}\
  \bibnamefont {Arizmendi-Carvajal}},\ }\href {\doibase
  10.1088/1751-8113/46/11/115006} {\bibfield  {journal} {\bibinfo  {journal}
  {Journal of Physics A: Mathematical and Theoretical}\ }\textbf {\bibinfo
  {volume} {46}},\ \bibinfo {pages} {115006} (\bibinfo {year}
  {2013})}\BibitemShut {NoStop}%
\bibitem [{\citenamefont {Dettori}\ \emph {et~al.}(2016)\citenamefont
  {Dettori}, \citenamefont {Melis}, \citenamefont {Rurali},\ and\ \citenamefont
  {Colombo}}]{Dettori2016}%
  \BibitemOpen
  \bibfield  {author} {\bibinfo {author} {\bibfnamefont {R.}~\bibnamefont
  {Dettori}}, \bibinfo {author} {\bibfnamefont {C.}~\bibnamefont {Melis}},
  \bibinfo {author} {\bibfnamefont {R.}~\bibnamefont {Rurali}}, \ and\ \bibinfo
  {author} {\bibfnamefont {L.}~\bibnamefont {Colombo}},\ }\href {\doibase
  10.1063/1.4953142} {\bibfield  {journal} {\bibinfo  {journal} {Journal of
  Applied Physics}\ }\textbf {\bibinfo {volume} {119}},\ \bibinfo {pages}
  {215102} (\bibinfo {year} {2016})}\BibitemShut {NoStop}%
\bibitem [{\citenamefont {Pereira}(2010)}]{Pereira2010}%
  \BibitemOpen
  \bibfield  {author} {\bibinfo {author} {\bibfnamefont {E.}~\bibnamefont
  {Pereira}},\ }\href {\doibase 10.1103/PhysRevE.82.040101} {\bibfield
  {journal} {\bibinfo  {journal} {Phys. Rev. E}\ }\textbf {\bibinfo {volume}
  {82}},\ \bibinfo {pages} {040101} (\bibinfo {year} {2010})}\BibitemShut
  {NoStop}%
\bibitem [{\citenamefont {Pereira}(2011)}]{Pereira2011}%
  \BibitemOpen
  \bibfield  {author} {\bibinfo {author} {\bibfnamefont {E.}~\bibnamefont
  {Pereira}},\ }\href {\doibase 10.1103/PhysRevE.83.031106} {\bibfield
  {journal} {\bibinfo  {journal} {Phys. Rev. E}\ }\textbf {\bibinfo {volume}
  {83}},\ \bibinfo {pages} {031106} (\bibinfo {year} {2011})}\BibitemShut
  {NoStop}%
\bibitem [{\citenamefont {{\'{A}}vila}\ and\ \citenamefont
  {Pereira}(2013)}]{Avila2013}%
  \BibitemOpen
  \bibfield  {author} {\bibinfo {author} {\bibfnamefont {R.~R.}\ \bibnamefont
  {{\'{A}}vila}}\ and\ \bibinfo {author} {\bibfnamefont {E.}~\bibnamefont
  {Pereira}},\ }\href {\doibase 10.1088/1751-8113/46/5/055002} {\bibfield
  {journal} {\bibinfo  {journal} {Journal of Physics A: Mathematical and
  Theoretical}\ }\textbf {\bibinfo {volume} {46}},\ \bibinfo {pages} {055002}
  (\bibinfo {year} {2013})}\BibitemShut {NoStop}%
\bibitem [{\citenamefont {Bagchi}(2017)}]{Bagchi2017}%
  \BibitemOpen
  \bibfield  {author} {\bibinfo {author} {\bibfnamefont {D.}~\bibnamefont
  {Bagchi}},\ }\href {\doibase 10.1103/PhysRevE.95.032102} {\bibfield
  {journal} {\bibinfo  {journal} {Phys. Rev. E}\ }\textbf {\bibinfo {volume}
  {95}},\ \bibinfo {pages} {032102} (\bibinfo {year} {2017})}\BibitemShut
  {NoStop}%
\bibitem [{\citenamefont {Pereira}\ and\ \citenamefont
  {\'Avila}(2013)}]{Pereira2013}%
  \BibitemOpen
  \bibfield  {author} {\bibinfo {author} {\bibfnamefont {E.}~\bibnamefont
  {Pereira}}\ and\ \bibinfo {author} {\bibfnamefont {R.~R.}\ \bibnamefont
  {\'Avila}},\ }\href {\doibase 10.1103/PhysRevE.88.032139} {\bibfield
  {journal} {\bibinfo  {journal} {Phys. Rev. E}\ }\textbf {\bibinfo {volume}
  {88}},\ \bibinfo {pages} {032139} (\bibinfo {year} {2013})}\BibitemShut
  {NoStop}%
\bibitem [{\citenamefont {Freitas}\ \emph {et~al.}(2016)\citenamefont
  {Freitas}, \citenamefont {Martinez},\ and\ \citenamefont
  {Paz}}]{Freitas2016}%
  \BibitemOpen
  \bibfield  {author} {\bibinfo {author} {\bibfnamefont {N.}~\bibnamefont
  {Freitas}}, \bibinfo {author} {\bibfnamefont {E.~A.}\ \bibnamefont
  {Martinez}}, \ and\ \bibinfo {author} {\bibfnamefont {J.~P.}\ \bibnamefont
  {Paz}},\ }\href {http://stacks.iop.org/1402-4896/91/i=1/a=013007} {\bibfield
  {journal} {\bibinfo  {journal} {Physica Scripta}\ }\textbf {\bibinfo {volume}
  {91}},\ \bibinfo {pages} {013007} (\bibinfo {year} {2016})}\BibitemShut
  {NoStop}%
\bibitem [{\citenamefont {Ruiz}\ \emph {et~al.}(2014)\citenamefont {Ruiz},
  \citenamefont {Alonso}, \citenamefont {Plenio},\ and\ \citenamefont {del
  Campo}}]{Ruiz2014}%
  \BibitemOpen
  \bibfield  {author} {\bibinfo {author} {\bibfnamefont {A.}~\bibnamefont
  {Ruiz}}, \bibinfo {author} {\bibfnamefont {D.}~\bibnamefont {Alonso}},
  \bibinfo {author} {\bibfnamefont {M.~B.}\ \bibnamefont {Plenio}}, \ and\
  \bibinfo {author} {\bibfnamefont {A.}~\bibnamefont {del Campo}},\ }\href
  {\doibase 10.1103/PhysRevB.89.214305} {\bibfield  {journal} {\bibinfo
  {journal} {Phys. Rev. B}\ }\textbf {\bibinfo {volume} {89}},\ \bibinfo
  {pages} {214305} (\bibinfo {year} {2014})}\BibitemShut {NoStop}%
\bibitem [{\citenamefont {Pruttivarasin}\ \emph {et~al.}(2011)\citenamefont
  {Pruttivarasin}, \citenamefont {Ramm}, \citenamefont {Talukdar},
  \citenamefont {Kreuter},\ and\ \citenamefont
  {H{\"a}ffner}}]{Pruttivarasin2011}%
  \BibitemOpen
  \bibfield  {author} {\bibinfo {author} {\bibfnamefont {T.}~\bibnamefont
  {Pruttivarasin}}, \bibinfo {author} {\bibfnamefont {M.}~\bibnamefont {Ramm}},
  \bibinfo {author} {\bibfnamefont {I.}~\bibnamefont {Talukdar}}, \bibinfo
  {author} {\bibfnamefont {A.}~\bibnamefont {Kreuter}}, \ and\ \bibinfo
  {author} {\bibfnamefont {H.}~\bibnamefont {H{\"a}ffner}},\ }\href {\doibase
  10.1088/1367-2630/13/7/075012} {\bibfield  {journal} {\bibinfo  {journal}
  {New Journal of Physics}\ }\textbf {\bibinfo {volume} {13}},\ \bibinfo
  {pages} {075012} (\bibinfo {year} {2011})}\BibitemShut {NoStop}%
\bibitem [{\citenamefont {Ramm}\ \emph {et~al.}(2014)\citenamefont {Ramm},
  \citenamefont {Pruttivarasin},\ and\ \citenamefont {H{\"a}ffner}}]{Ramm2014}%
  \BibitemOpen
  \bibfield  {author} {\bibinfo {author} {\bibfnamefont {M.}~\bibnamefont
  {Ramm}}, \bibinfo {author} {\bibfnamefont {T.}~\bibnamefont {Pruttivarasin}},
  \ and\ \bibinfo {author} {\bibfnamefont {H.}~\bibnamefont {H{\"a}ffner}},\
  }\href {http://stacks.iop.org/1367-2630/16/i=6/a=063062} {\bibfield
  {journal} {\bibinfo  {journal} {New Journal of Physics}\ }\textbf {\bibinfo
  {volume} {16}},\ \bibinfo {pages} {063062} (\bibinfo {year}
  {2014})}\BibitemShut {NoStop}%
\bibitem [{\citenamefont {Cirac}\ and\ \citenamefont
  {Zoller}(2000)}]{Cirac2000}%
  \BibitemOpen
  \bibfield  {author} {\bibinfo {author} {\bibfnamefont {J.~I.}\ \bibnamefont
  {Cirac}}\ and\ \bibinfo {author} {\bibfnamefont {P.}~\bibnamefont {Zoller}},\
  }\href {https://doi.org/10.1038/35007021} {\bibfield  {journal} {\bibinfo
  {journal} {Nature}\ }\textbf {\bibinfo {volume} {404}},\ \bibinfo {pages}
  {579 EP } (\bibinfo {year} {2000})}\BibitemShut {NoStop}%
\bibitem [{\citenamefont {Krauth}\ \emph {et~al.}(2014)\citenamefont {Krauth},
  \citenamefont {Alonso},\ and\ \citenamefont {Home}}]{Krauth2014}%
  \BibitemOpen
  \bibfield  {author} {\bibinfo {author} {\bibfnamefont {F.}~\bibnamefont
  {Krauth}}, \bibinfo {author} {\bibfnamefont {J.}~\bibnamefont {Alonso}}, \
  and\ \bibinfo {author} {\bibfnamefont {J.}~\bibnamefont {Home}},\ }\href@noop
  {} {\bibfield  {journal} {\bibinfo  {journal} {Journal of Physics B: Atomic,
  Molecular and Optical Physics}\ }\textbf {\bibinfo {volume} {48}},\ \bibinfo
  {pages} {015001} (\bibinfo {year} {2014})}\BibitemShut {NoStop}%
\bibitem [{\citenamefont {Schmied}\ \emph {et~al.}(2009)\citenamefont
  {Schmied}, \citenamefont {Wesenberg},\ and\ \citenamefont
  {Leibfried}}]{Schmied2009}%
  \BibitemOpen
  \bibfield  {author} {\bibinfo {author} {\bibfnamefont {R.}~\bibnamefont
  {Schmied}}, \bibinfo {author} {\bibfnamefont {J.~H.}\ \bibnamefont
  {Wesenberg}}, \ and\ \bibinfo {author} {\bibfnamefont {D.}~\bibnamefont
  {Leibfried}},\ }\href {\doibase 10.1103/PhysRevLett.102.233002} {\bibfield
  {journal} {\bibinfo  {journal} {Phys. Rev. Lett.}\ }\textbf {\bibinfo
  {volume} {102}},\ \bibinfo {pages} {233002} (\bibinfo {year}
  {2009})}\BibitemShut {NoStop}%
\bibitem [{\citenamefont {Lepri}\ \emph {et~al.}(2003)\citenamefont {Lepri},
  \citenamefont {Livi},\ and\ \citenamefont {Politi}}]{Lepri2003}%
  \BibitemOpen
  \bibfield  {author} {\bibinfo {author} {\bibfnamefont {S.}~\bibnamefont
  {Lepri}}, \bibinfo {author} {\bibfnamefont {R.}~\bibnamefont {Livi}}, \ and\
  \bibinfo {author} {\bibfnamefont {A.}~\bibnamefont {Politi}},\ }\href
  {\doibase https://doi.org/10.1016/S0370-1573(02)00558-6} {\bibfield
  {journal} {\bibinfo  {journal} {Physics Reports}\ }\textbf {\bibinfo {volume}
  {377}},\ \bibinfo {pages} {1 } (\bibinfo {year} {2003})}\BibitemShut
  {NoStop}%
\bibitem [{\citenamefont {Dhar}(2008)}]{Dhar2018}%
  \BibitemOpen
  \bibfield  {author} {\bibinfo {author} {\bibfnamefont {A.}~\bibnamefont
  {Dhar}},\ }\href {\doibase 10.1080/00018730802538522} {\bibfield  {journal}
  {\bibinfo  {journal} {Advances in Physics}\ }\textbf {\bibinfo {volume}
  {57}},\ \bibinfo {pages} {457} (\bibinfo {year} {2008})}\BibitemShut
  {NoStop}%
\bibitem [{\citenamefont {Toral}\ and\ \citenamefont
  {Colet}(2014)}]{Toral2014}%
  \BibitemOpen
  \bibfield  {author} {\bibinfo {author} {\bibfnamefont {R.}~\bibnamefont
  {Toral}}\ and\ \bibinfo {author} {\bibfnamefont {P.}~\bibnamefont {Colet}},\
  }\href@noop {} {\emph {\bibinfo {title} {Stochastic numerical methods: an
  introduction for students and scientists}}}\ (\bibinfo  {publisher} {John
  Wiley \& Sons},\ \bibinfo {year} {2014})\BibitemShut {NoStop}%
\bibitem [{\citenamefont {Arimondo}\ \emph {et~al.}(1993)\citenamefont
  {Arimondo}, \citenamefont {Phillips},\ and\ \citenamefont
  {Strumia}}]{Arimondo1993}%
  \BibitemOpen
  \bibfield  {author} {\bibinfo {author} {\bibfnamefont {E.}~\bibnamefont
  {Arimondo}}, \bibinfo {author} {\bibfnamefont {W.}~\bibnamefont {Phillips}},
  \ and\ \bibinfo {author} {\bibfnamefont {F.}~\bibnamefont {Strumia}},\
  }\href@noop {} {\emph {\bibinfo {title} {Laser Manipulation of Atoms and
  Ions}}},\ Enrico Fermi International School of Physics\ (\bibinfo
  {publisher} {Elsevier Science},\ \bibinfo {year} {1993})\BibitemShut
  {NoStop}%
\bibitem [{\citenamefont {Metcalf}\ and\ \citenamefont {Van~der
  Straten}(1999)}]{Metcalf1999}%
  \BibitemOpen
  \bibfield  {author} {\bibinfo {author} {\bibfnamefont {H.}~\bibnamefont
  {Metcalf}}\ and\ \bibinfo {author} {\bibfnamefont {P.}~\bibnamefont {Van~der
  Straten}},\ }\href@noop {} {\emph {\bibinfo {title} {Laser Cooling and
  Trapping}}},\ Graduate texts in contemporary physics\ (\bibinfo  {publisher}
  {Springer},\ \bibinfo {year} {1999})\BibitemShut {NoStop}%
\bibitem [{\citenamefont {Metcalf}\ and\ \citenamefont {van~der
  Straten}(2003)}]{Metcalf2003}%
  \BibitemOpen
  \bibfield  {author} {\bibinfo {author} {\bibfnamefont {H.~J.}\ \bibnamefont
  {Metcalf}}\ and\ \bibinfo {author} {\bibfnamefont {P.}~\bibnamefont {van~der
  Straten}},\ }\href {\doibase 10.1364/JOSAB.20.000887} {\bibfield  {journal}
  {\bibinfo  {journal} {J. Opt. Soc. Am. B}\ }\textbf {\bibinfo {volume}
  {20}},\ \bibinfo {pages} {887} (\bibinfo {year} {2003})}\BibitemShut
  {NoStop}%
\bibitem [{\citenamefont {Cohen-Tannoudji}(1990)}]{Cohen1990}%
  \BibitemOpen
  \bibfield  {author} {\bibinfo {author} {\bibfnamefont {C.}~\bibnamefont
  {Cohen-Tannoudji}},\ }\href@noop {} {\bibfield  {journal} {\bibinfo
  {journal} {Fundamental systems in quantum optics}\ ,\ \bibinfo {pages} {1}}
  (\bibinfo {year} {1990})}\BibitemShut {NoStop}%
\bibitem [{\citenamefont {{Chee Kong}}(2010)}]{Chee2010}%
  \BibitemOpen
  \bibfield  {author} {\bibinfo {author} {\bibfnamefont {L.}~\bibnamefont
  {{Chee Kong}}},\ }\href@noop {} {\bibfield  {journal} {\bibinfo  {journal}
  {arXiv e-prints}\ ,\ \bibinfo {eid} {arXiv:1008.0491}} (\bibinfo {year}
  {2010})}\BibitemShut {NoStop}%
\bibitem [{\citenamefont {{Novikov}}(1965)}]{Novikov1965}%
  \BibitemOpen
  \bibfield  {author} {\bibinfo {author} {\bibfnamefont {E.~A.}\ \bibnamefont
  {{Novikov}}},\ }\href@noop {} {\bibfield  {journal} {\bibinfo  {journal}
  {Soviet Journal of Experimental and Theoretical Physics}\ }\textbf {\bibinfo
  {volume} {20}},\ \bibinfo {pages} {1290} (\bibinfo {year}
  {1965})}\BibitemShut {NoStop}%
\bibitem [{\citenamefont {Ma}\ and\ \citenamefont {Dudarev}(2011)}]{Ma2011}%
  \BibitemOpen
  \bibfield  {author} {\bibinfo {author} {\bibfnamefont {P.-W.}\ \bibnamefont
  {Ma}}\ and\ \bibinfo {author} {\bibfnamefont {S.~L.}\ \bibnamefont
  {Dudarev}},\ }\href {\doibase 10.1103/PhysRevB.83.134418} {\bibfield
  {journal} {\bibinfo  {journal} {Phys. Rev. B}\ }\textbf {\bibinfo {volume}
  {83}},\ \bibinfo {pages} {134418} (\bibinfo {year} {2011})}\BibitemShut
  {NoStop}%
\bibitem [{\citenamefont {S\"arkk\"a}\ and\ \citenamefont
  {Solin}(2019)}]{Sarkka2019}%
  \BibitemOpen
  \bibfield  {author} {\bibinfo {author} {\bibfnamefont {S.}~\bibnamefont
  {S\"arkk\"a}}\ and\ \bibinfo {author} {\bibfnamefont {A.}~\bibnamefont
  {Solin}},\ }\href@noop {} {\emph {\bibinfo {title} {Applied Stochastic
  Differential Equations}}}\ (\bibinfo  {publisher} {Cambridge University
  Press},\ \bibinfo {year} {2019})\BibitemShut {NoStop}%
\bibitem [{\citenamefont {James}(1998)}]{James1998}%
  \BibitemOpen
  \bibfield  {author} {\bibinfo {author} {\bibfnamefont {D.~F.}\ \bibnamefont
  {James}},\ }\href@noop {} {\bibfield  {journal} {\bibinfo  {journal} {Applied
  Physics B: Lasers and Optics}\ }\textbf {\bibinfo {volume} {66}},\ \bibinfo
  {pages} {181} (\bibinfo {year} {1998})}\BibitemShut {NoStop}%
\bibitem [{\citenamefont {Bezanson}\ \emph {et~al.}(2012)\citenamefont
  {Bezanson}, \citenamefont {Karpinski}, \citenamefont {Shah},\ and\
  \citenamefont {Edelman}}]{Bezanson2012}%
  \BibitemOpen
  \bibfield  {author} {\bibinfo {author} {\bibfnamefont {J.}~\bibnamefont
  {Bezanson}}, \bibinfo {author} {\bibfnamefont {S.}~\bibnamefont {Karpinski}},
  \bibinfo {author} {\bibfnamefont {V.~B.}\ \bibnamefont {Shah}}, \ and\
  \bibinfo {author} {\bibfnamefont {A.}~\bibnamefont {Edelman}},\ }\href
  {https://arxiv.org/abs/1209.5145} {\bibfield  {journal} {\bibinfo  {journal}
  {arXiv:1209.5145}\ } (\bibinfo {year} {2012})}\BibitemShut {NoStop}%
\bibitem [{\citenamefont {Bezanson}\ \emph {et~al.}(2017)\citenamefont
  {Bezanson}, \citenamefont {Edelman}, \citenamefont {Karpinski},\ and\
  \citenamefont {Shah}}]{Bezanson2017}%
  \BibitemOpen
  \bibfield  {author} {\bibinfo {author} {\bibfnamefont {J.}~\bibnamefont
  {Bezanson}}, \bibinfo {author} {\bibfnamefont {A.}~\bibnamefont {Edelman}},
  \bibinfo {author} {\bibfnamefont {S.}~\bibnamefont {Karpinski}}, \ and\
  \bibinfo {author} {\bibfnamefont {V.}~\bibnamefont {Shah}},\ }\href {\doibase
  10.1137/141000671} {\bibfield  {journal} {\bibinfo  {journal} {SIAM Review}\
  }\textbf {\bibinfo {volume} {59}},\ \bibinfo {pages} {65} (\bibinfo {year}
  {2017})}\BibitemShut {NoStop}%
\bibitem [{\citenamefont {Rackauckas}\ and\ \citenamefont
  {Nie}(2017)}]{Rackauckas2017}%
  \BibitemOpen
  \bibfield  {author} {\bibinfo {author} {\bibfnamefont {C.}~\bibnamefont
  {Rackauckas}}\ and\ \bibinfo {author} {\bibfnamefont {Q.}~\bibnamefont
  {Nie}},\ }\href {http://aimsciences.org//article/doi/10.3934/dcdsb.2017133}
  {\bibfield  {journal} {\bibinfo  {journal} {Discrete and continuous dynamical
  systems. Series B}\ }\textbf {\bibinfo {volume} {22}},\ \bibinfo {pages}
  {2731} (\bibinfo {year} {2017})}\BibitemShut {NoStop}%
\bibitem [{\citenamefont {Leupold}(2015)}]{Leupold2015}%
  \BibitemOpen
  \bibfield  {author} {\bibinfo {author} {\bibfnamefont {F.~M.}\ \bibnamefont
  {Leupold}},\ }\emph {\bibinfo {title} {Bang-bang Control of a Trapped-Ion
  Oscillator}},\ \href {\doibase 10.3929/ethz-a-010616440} {Ph.D. thesis},\
  \bibinfo  {school} {ETH Zurich} (\bibinfo {year} {2015})\BibitemShut
  {NoStop}%
\bibitem [{\citenamefont {Lo}(2015)}]{Lo2015}%
  \BibitemOpen
  \bibfield  {author} {\bibinfo {author} {\bibfnamefont {H.-Y.}\ \bibnamefont
  {Lo}},\ }\emph {\bibinfo {title} {Creation of Squeezed Schr\"odinger's Cat
  States in a Mixed-Species Ion Trap}},\ \href {\doibase
  10.3929/ethz-a-010592649} {Ph.D. thesis},\ \bibinfo  {school} {ETH Zurich}
  (\bibinfo {year} {2015})\BibitemShut {NoStop}%
\bibitem [{\citenamefont {Enderlein}\ \emph {et~al.}(2012)\citenamefont
  {Enderlein}, \citenamefont {Huber}, \citenamefont {Schneider},\ and\
  \citenamefont {Schaetz}}]{Enderlein2012}%
  \BibitemOpen
  \bibfield  {author} {\bibinfo {author} {\bibfnamefont {M.}~\bibnamefont
  {Enderlein}}, \bibinfo {author} {\bibfnamefont {T.}~\bibnamefont {Huber}},
  \bibinfo {author} {\bibfnamefont {C.}~\bibnamefont {Schneider}}, \ and\
  \bibinfo {author} {\bibfnamefont {T.}~\bibnamefont {Schaetz}},\ }\href
  {\doibase 10.1103/PhysRevLett.109.233004} {\bibfield  {journal} {\bibinfo
  {journal} {Phys. Rev. Lett.}\ }\textbf {\bibinfo {volume} {109}},\ \bibinfo
  {pages} {233004} (\bibinfo {year} {2012})}\BibitemShut {NoStop}%
\bibitem [{\citenamefont {Bermudez}\ \emph {et~al.}(2013)\citenamefont
  {Bermudez}, \citenamefont {Bruderer},\ and\ \citenamefont
  {Plenio}}]{Bermudez2013}%
  \BibitemOpen
  \bibfield  {author} {\bibinfo {author} {\bibfnamefont {A.}~\bibnamefont
  {Bermudez}}, \bibinfo {author} {\bibfnamefont {M.}~\bibnamefont {Bruderer}},
  \ and\ \bibinfo {author} {\bibfnamefont {M.~B.}\ \bibnamefont {Plenio}},\
  }\href {\doibase 10.1103/PhysRevLett.111.040601} {\bibfield  {journal}
  {\bibinfo  {journal} {Phys. Rev. Lett.}\ }\textbf {\bibinfo {volume} {111}},\
  \bibinfo {pages} {040601} (\bibinfo {year} {2013})}\BibitemShut {NoStop}%
\bibitem [{\citenamefont {Schneider}\ \emph {et~al.}(2010)\citenamefont
  {Schneider}, \citenamefont {Enderlein}, \citenamefont {Huber},\ and\
  \citenamefont {Schaetz}}]{Schneider2010}%
  \BibitemOpen
  \bibfield  {author} {\bibinfo {author} {\bibfnamefont {C.}~\bibnamefont
  {Schneider}}, \bibinfo {author} {\bibfnamefont {M.}~\bibnamefont
  {Enderlein}}, \bibinfo {author} {\bibfnamefont {T.}~\bibnamefont {Huber}}, \
  and\ \bibinfo {author} {\bibfnamefont {T.}~\bibnamefont {Schaetz}},\ }\href
  {https://doi.org/10.1038/nphoton.2010.236} {\bibfield  {journal} {\bibinfo
  {journal} {Nature Photonics}\ }\textbf {\bibinfo {volume} {4}},\ \bibinfo
  {pages} {772 EP } (\bibinfo {year} {2010})}\BibitemShut {NoStop}%
\bibitem [{\citenamefont {Riera-Campeny}\ \emph {et~al.}(2019)\citenamefont
  {Riera-Campeny}, \citenamefont {Mehboudi}, \citenamefont {Pons},\ and\
  \citenamefont {Sanpera}}]{Riera-Campeny2018}%
  \BibitemOpen
  \bibfield  {author} {\bibinfo {author} {\bibfnamefont {A.}~\bibnamefont
  {Riera-Campeny}}, \bibinfo {author} {\bibfnamefont {M.}~\bibnamefont
  {Mehboudi}}, \bibinfo {author} {\bibfnamefont {M.}~\bibnamefont {Pons}}, \
  and\ \bibinfo {author} {\bibfnamefont {A.}~\bibnamefont {Sanpera}},\ }\href
  {\doibase 10.1103/PhysRevE.99.032126} {\bibfield  {journal} {\bibinfo
  {journal} {Phys. Rev. E}\ }\textbf {\bibinfo {volume} {99}},\ \bibinfo
  {pages} {032126} (\bibinfo {year} {2019})}\BibitemShut {NoStop}%
\end{thebibliography}%

\end{document}